\documentclass[reprint,aps,prb,showpacs]{revtex4-1}
\usepackage{color,graphicx,calc,url}
\usepackage{dcolumn}
\usepackage{bm}
\usepackage{amssymb}
\usepackage{amsmath}
\usepackage{wasysym}
\usepackage{natbib}
\usepackage{mathrsfs}

\begin{document}
\raggedbottom
\title{Enhancement of thermovoltage and tunnel magneto-Seebeck effect in CoFeB based magnetic tunnel junctions by variation of the MgAl$_2$O$_4$ and MgO barrier thickness}
\author{Torsten Huebner,$^1$ Ulrike Martens,$^2$ Jakob Walowski,$^2$ Alexander Boehnke,$^1$ Jan Krieft,$^1$ Christian Heiliger,$^{3}$ Andy Thomas,$^4$  G\"unter Reiss,$^1$ Timo Kuschel,$^{1,5}$ Markus M\"unzenberg$^2$ \email{Electronic mail: thuebner@physik.uni-bielefeld.de}}
\affiliation{$^1$Center for Spinelectronic Materials and Devices, Department of Physics, Bielefeld University, Universit\"atsstra\ss e 25, 33615 Bielefeld, Germany\\
$^2$Institut f\"ur Physik, Greifswald University, Felix-Hausdorff-Strasse 6, 17489 Greifswald, Germany\\
$^3$I. Physikalisches Institut, Justus Liebig University Giessen, Heinrich-Buff-Ring 16, 35392 Giessen, Germany\\
$^4$Institute for Metallic Materials, IFW Dresden, Helmholtzstra\ss e 20, 01069 Dresden, Germany\\
$^5$Physics of Nanodevices, Zernike Institute for Advanced Materials, University of Groningen, Nijenborgh 4, 9747 AG Groningen, The Netherlands}
\date{\today}


\begin{abstract}
We investigate the influence of the barrier thickness of Co$_{40}$Fe$_{40}$B$_{20}$ based magnetic tunnel junctions on the laser-induced tunnel magneto-Seebeck effect. 
Varying the barrier thickness from 1\,nm to 3\,nm, we find a distinct maximum in the tunnel magneto-Seebeck effect for 2.6\,nm barrier thickness. This maximum is independently measured for two barrier materials, namely MgAl$_2$O$_4$ and MgO. Additionally, samples with an MgAl$_2$O$_4$ barrier exhibit a high thermovoltage of more than 350\,$\mu$V in comparison to 90\,$\mu$V for the MTJs with MgO barrier when heated with the maximum laser power of 150\,mW. Our results allow for the fabrication of improved stacks when dealing with temperature differences across magnetic tunnel junctions for future applications in spin caloritronics, the emerging research field that combines spintronics and themoelectrics. 
\end{abstract}

\maketitle
\section{Introduction}

In recent years, the combination of the spintronic \textit{magnetic tunnel junction} (MTJ) and a temperature gradient was intensively studied. Since these experiments combine spin, charge and heat driven currents, they are prominent examples for the hot topic of spin caloritronics \cite{Bauer}. At first, the \textit{tunnel magneto-Seebeck} (TMS) effect was predicted \cite{czerner} and measured with two different techniques \cite{walter,Liebing1}. Later on, the experimentally even more challenging \textit{tunnel magneto-Peltier effect}, which is reciprocal to the TMS effect, was observed as well \cite{magnetopeltier}. 

Subsequent studies focused on the increase of effect sizes, film quality and the overcoming of experimental challenges. In particular, a giant TMS ratio of -3000\,\% was found when applying an additional bias voltage across the MTJ \cite{Boehnke2}, a significant improvement of the TMS ratio was obtained with the usage of half-metallic electrodes from ferromagnetic Heusler compounds such as Co$_2$FeAl or Co$_2$FeSi \cite{Boehnke3} and parasitic effects originating from semiconducting substrates were clarified \cite{Boehnke1}. Additionally, in a preceding publication \cite{huebner}, we compared the laser-induced TMS with the method of the intrinsic TMS, which uses a symmetry analysis of the tunneling current with respect to the applied voltage. The Brinkman model \cite{brinkman} offered an alternative way to explain the symmetric contribution previously associated with the intrinsic TMS. Thus, we concluded that it is not possible to explicitly observe an intrinsic TMS. 

Up to now, theoretical works focused only on 6 or 10 atomic layers of barrier thickness, respectively, and on the electrode/barrier interface, which hugely influences not only the TMR but also the TMS effect \cite{czerner2,heiliger}. Fe-Co/MgO is often used as a model system within these studies due to the large computational effort that is necessary, e.g., to model the TMS for materials with a more complex crystal structure. Furthermore, Fe-Co/MgO exhibits coherent tunneling of the electrons via $\Delta_1$ states and, thus, ensures high TMR ratios needed for applications. A combination of an additionally applied temperature gradient and the continuing improvement of Seebeck voltages and TMS ratios may support the development of 'milivolt switches' \cite{milivolt} based, for example, on the \textit{thermal spin-transfer torque} \cite{TSTT}. 

Previous TMS measurements concentrated on the established MTJ system of Co-Fe(CoFeB)/MgO with a standard barrier thickness of around 2\,nm. Therefore, we investigate the system of CoFeB and MgAl$_2$O$_4$ (MAO) with different barrier thicknesses and junction sizes in order to maximize the TMS effect. Theoretically, MAO exhibits an advantageous lattice mismatch (1\,\%) with standard ferromagnetic electrodes such as Fe, CoFe or CoFeB when compared to MgO ((3-5)\,\%) \cite{miura}. As a barrier, MAO also enables coherent tunneling via the $\Delta_1$ symmetry filter effect \cite{meff}. So far, experimental results of the TMR effect in MAO MTJs fall short in comparison to MgO MTJs \cite{Sukegawa1,tao2,scheike}, but, for example, magnetization switching by \textit{spin-transfer torque} has been demonstrated \cite{Sukegawa2}. Additionally, by growing MAO barriers via \textit{molecular beam epitaxy}, MgAl$_2$O$_x$ double-barrier MTJs exhibit almost no lattice mismatch between electrode and barrier showing pronounced resonant tunneling features in quantum well structures \cite{tao}. As a direct comparison with recent experiments and theoretical predictions, we compare our results for MAO barriers with CoFeB/MgO MTJs.

This paper is organized as follows: Sec. \ref{prep} starts with the sample deposition and preparation, followed by Sec. \ref{results}, which is split into three subsections. Here, Sec. \ref{results1} deals with the results of the TMR and TMS measurements, Sec. \ref{results2} with the results of the I/V curves and Sec. \ref{results3} with the thermovoltage and COMSOL evaluation. Section \ref{conclusion} concludes this paper. 
 
\section{Sample deposition and preparation}\label{prep}

We prepared different sample series in order to give a detailed overview concerning reproducibility and comparability. The MAO and MgO MTJs are sputtered in a Leybold Vakuum GmbH CLAB 600 cluster tool at a base pressure of less than $5\cdot10^{-7}$\,mbar. This system allows the deposition of several samples without exposing them to ambient conditions in between sputtering processes. The whole stack of all series is composed of a bottom contact of Ta 10/Ru 30/Ta 5/Ru 5, a tunnel junction of MnIr 10/Co$_{40}$Fe$_{40}$B$_{20}$ 2.5/barrier/Co$_{40}$Fe$_{40}$B$_{20}$ 2.5 and a top contact of Ta 5/Ru 30/Ta 5/Au 60 (numbers are nominal thicknesses in nm). The resulting sample series are summarized in Tab. \ref{sampleseries}.

To achieve the exchange biasing of the ferromagnetic electrode by the MnIr, the stacks are post annealed at 350\,$^{\circ}$C for one hour, followed by cooling in a magnetic field of 0.7\,T. \textit{Electron beam lithography} and \textit{ion beam etching} is used to pattern elliptical junctions of $0.5\pi$\,$\mu$m$^2$, $2\pi$\,$\mu$m$^2$ and $6\pi$\,$\mu$m$^2$ with the major axis being twice as large as the minor axis. Ta$_2$O$_5$ (120\,nm) is used as insulating material between individual MTJs and Au bond pads serve as electrical contacts and heat absorbers. More details can be found in preceding publications \cite{walter,Boehnke1,Boehnke2,huebner}. 
\begin{table}[bt]\centering
	\caption{Overview of different nominal barrier thicknesses of each series.}
	\label{sampleseries}
	%\vspace{0.1cm}
		\begin{tabular}{c| c}
		\hline\hline
			Series & Nominal barrier thickness (nm) \\
			\hline
			I (MAO) & 1.0 1.4 1.6 1.8$^\text{a}$ 2.0$^\text{a}$ 2.2 2.6 3.0\\
			II (MgO) & 1.2 1.5 1.8 1.9 2.0 2.3 2.6 2.9\\
			\hline \hline
		\end{tabular}
		\\
			\vspace{0.25cm}
		$^\text{a}$Samples were prepared independently of the rest of the series.\\
\end{table}

\section{Results}\label{results}

\subsection{TMR and TMS results}\label{results1}

\begin{figure}[bt]\centering
		\includegraphics{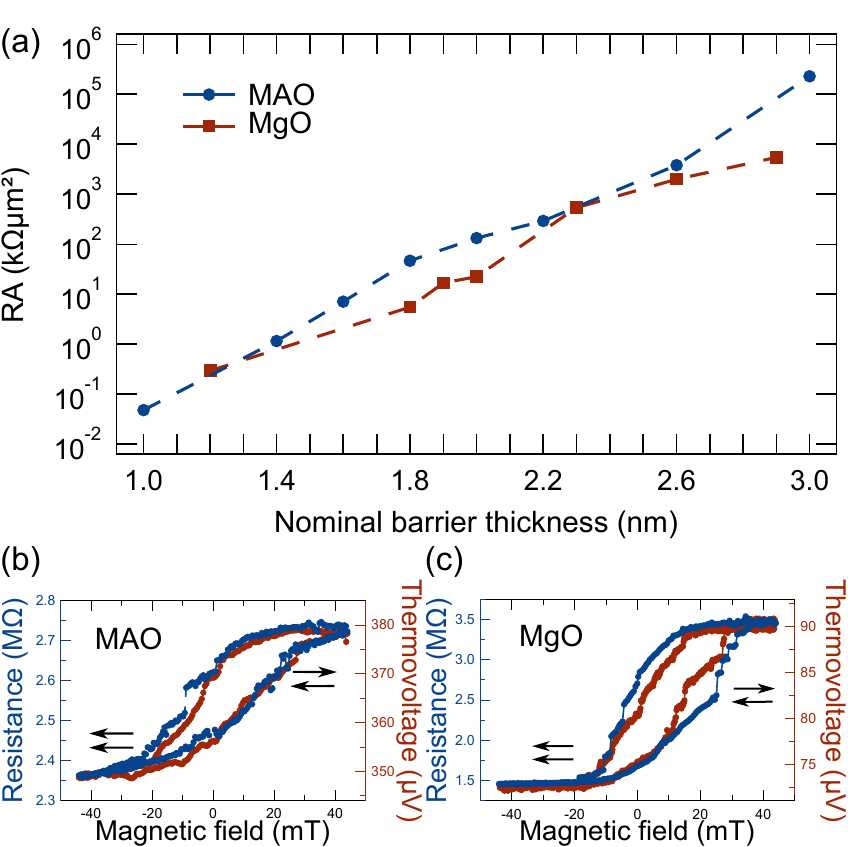}
	\caption{(a) Averaged RA products for the MAO MTJs (blue ircles) and the MgO MTJs (red squares) in the parallel state. (b,c) Exemplary minor loops with the highest switching ratios of the samples with MAO and MgO barrier, respectively. Both loops are measured at the smallest junction area of $0.5\pi$$\mu$m$^2$.}
	\label{fig:RA}
\end{figure}

Figure \ref{fig:RA}(a) shows the resistance area (RA) products of both series in dependence of the nominal barrier thicknesses in the parallel magnetization alignment. Please note that the error bars of the RA product are too small to be seen, indicating an overall homogenous sample quality of all series. As expected, the RA product increases exponentially with increasing nominal barrier thickness. 

In addition, the RA products of the independently prepared samples within series I fit very well together, indicating that the nominal is close to the real barrier thickness. Since the RA product is mostly governed by the barrier, it is noteworthy that the different barrier materials lead to comparable RA values between the series. Two minor loops of the samples with the highest TMS ratios are shown in Fig. \ref{fig:RA}(b,c). Here, the nominal barrier thickness is 2.6\,nm and the junction size is $0.5\pi$$\mu$m$^2$ in both cases. Despite the high resistance resulting from the thick barrier of 2.6\,nm, both MTJs show clear parallel and antiparallel states with the same switching behavior for TMS and TMR measurements. The extracted TMS (TMR) ratio amounts to 8\,\% (18\,\%) for MAO whereas it is 28\,\% (130\,\%) for MgO. The sample with MAO barrier shows a very high thermovoltage of around 375\,$\mu$V in contrast to around 80\,$\mu$V in case of an MgO barrier when using a laser power of 150\,mW.

\begin{figure}[bt]\centering
		\includegraphics{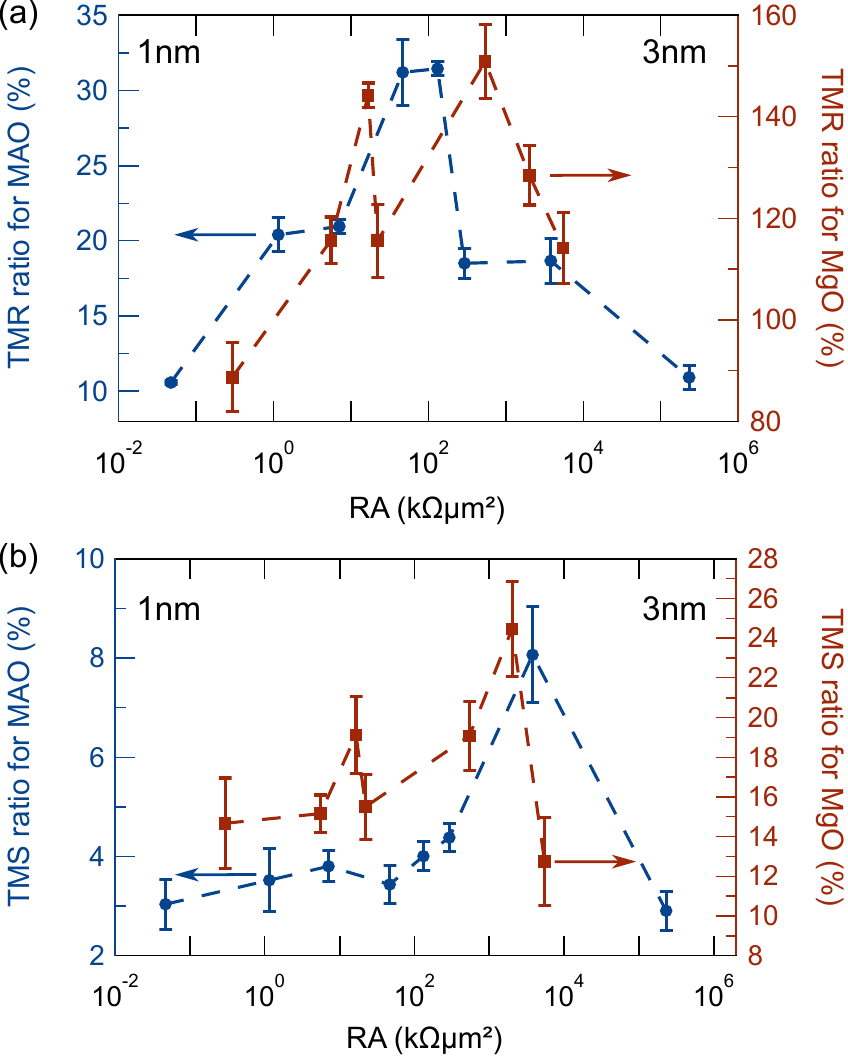}
	\caption{(a) Averaged TMR ratios of all measured elements with resulting error bars versus RA: MAO (blue circles) and MgO (red squares). (b) Averaged TMS ratios of all measured elements with resulting error bars versus RA: MAO (blue circles) and MgO (red squares).}
	\label{fig:TMR_TMS}
\end{figure}

Figure \ref{fig:TMR_TMS}(a) summarizes the results of the TMR measurements of series I and II in dependence of the RA product. For each barrier thickness several elements as well as different element areas are measured and averaged. Firstly, both barrier materials show TMR maximum values (MAO: 30\,\%, MgO: 150\,\%) around a nominal barrier thickness of 2\,nm (RA$_{\text{MAO}}\approx100\,\text{k}\Omega\mu\text{m}^2$, RA$_{\text{MgO}}\approx1000\,\text{k}\Omega\mu\text{m}^2$). Secondly, the series with MgO barrier exhibits a second peak of the TMR for a barrier thickness of 1.9\,nm (RA$=10\,\text{k}\Omega\mu\text{m}^2$). This peak might be directly related to the slightly increased RA product (c.f. Fig. \ref{fig:RA}(a)) in this region.

The dependence of the TMS ratio on the barrier thickness of both series is shown in Fig. \ref{fig:TMR_TMS}(b). Thin barriers of MAO exhibit a gradual increase of TMS ratios from 3\,\% to 4\,\%, while a distinct maximum is observed for a nominal barrier thickness of 2.6\,nm. Here, the TMS ratio doubles to 8\,\%. Furthermore, the TMS ratio of the MTJs with MgO barrier shows a similar behavior. It rises from 14\,\% to 19\,\% in case of thin barriers and shoots up to almost 28\,\% for a nominal barrier thickness of 2.6\,nm. In between, a local maximum is observable that directly corresponds to the position of the local TMR maximum. Usually, a direct correlation between TMR and TMS is not expected. For both barrier materials, the TMS peak is located around the same value of RA of some $10^3\,\text{k}\Omega\mu\text{m}^2$, which corresponds to a nominal barrier thickness of 2.6\,nm. Also, the TMS ratios of the samples prepared separately correspond well to the results of the rest of series I. 

In contrast to our experiments, theoretical calculations predict an increasing TMS ratio when going down from 10 monolayers (2\,\%) to 6 monolayers (10\,\%) of MgO \cite{heiliger}(1\,ML $\cong$ 2.1\,\text{\AA}). A reason for these opposite results might be a different interface structure of the electrode and the barrier, which is assumed to be perfectly ordered in the calculations.

\subsection{I/V measurements}\label{results2}

\begin{figure}[bt]\centering
		\includegraphics{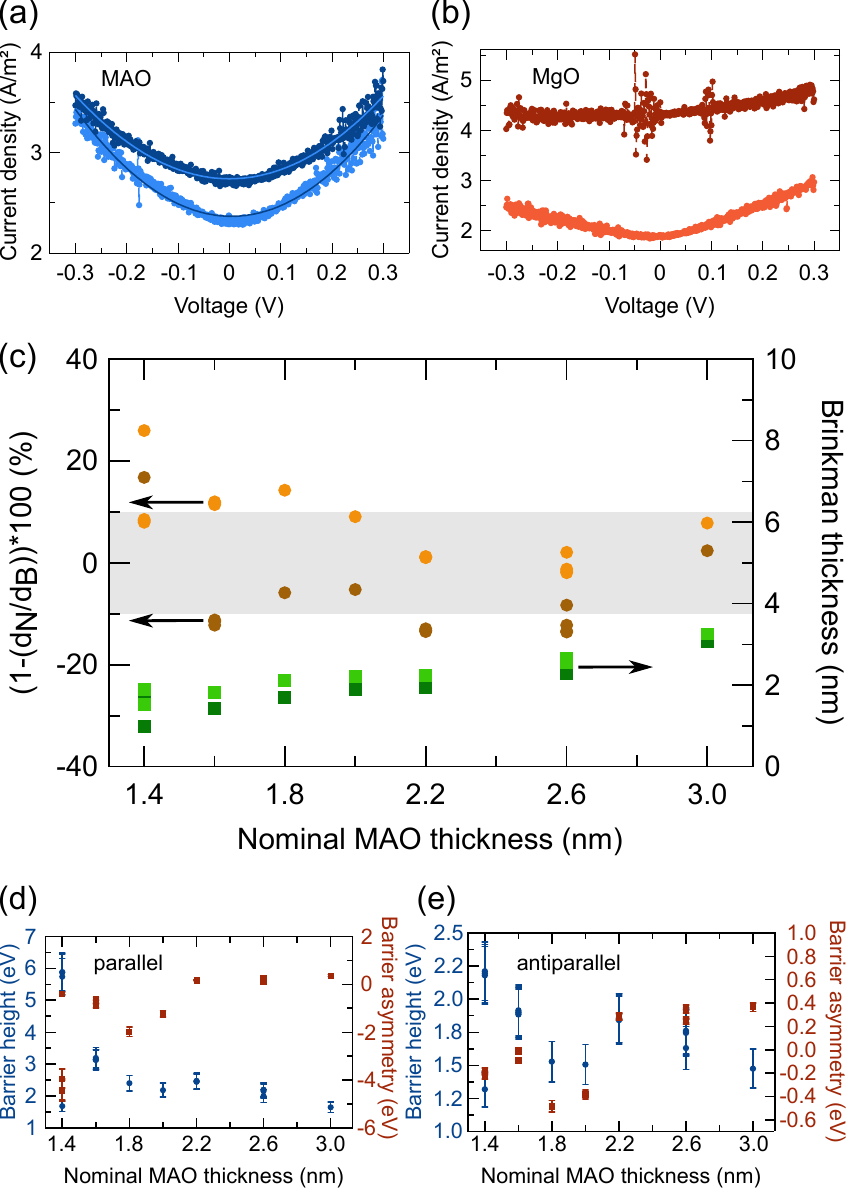}
	\caption{(a) dJ/dV data of the MTJ with MAO barrier (2.6\,nm, $0.5\pi$$\mu$m$^2$) with Brinkman fits that are used to extract barrier parameters (dark: parallel, light: antiparallel). (b) dJ/dV data of the MTJ with MgO barrier (2.6\,nm, $0.5\pi$$\mu$m$^2$) (dark: parallel, light: antiparallel). (c) Relative deviation of the nominal barrier thickness ($d_\text{N}$) and the calculated Brinkman barrier thickness ($d_\text{B}$). Dark (light) orange represents the results of the parallel (antiparallel) state. Please note that several element sizes are depicted that are partially overlapping in order to demonstrate the consistency of the method. The green squares are the extracted Brinkman thicknesses for each nominal MAO barrier thickness (dark: parallel, light: antiparallel). The gray area represents the typical error range of the Brinkman model of 10\,\%. (d,e) Barrier height (blue circles) and asymmetry (red squares) values for all measured elements and barrier thicknesses in the parallel (d) and antiparallel (e) state.}
	\label{fig:brink}
\end{figure}

Figures \ref{fig:brink}(a,b) show the dJ/dV (recalculated from I/V measurements) curves that are measured at the same elements as in Figs. \ref{fig:RA}(b,c). While the curves look similar in case of an MTJ with MAO barrier, they look very different in case of the MTJ with MgO barrier. This difference is due to the coherent tunneling of MgO based MTJs: The parallel curve is almost linear, while the antiparallel curve exhibits a pronounced kink around a bias voltage of 0\,V. Since the MAO MTJs exhibit a rather low TMR, no $\Delta_1$ symmetry filter effect and, thus, no coherent tunneling is present in the MTJs with MAO barrier. 

In order to further analyze the MTJs with MAO barrier, we use the Brinkman model, which allows to calculate the barrier height $\varphi$, the barrier asymmetry $\Delta\varphi$ and the barrier thickness $d_\text{B}$ from I/V measurements. A theoretical description of this model can be found in Ref. \onlinecite{brinkman}, while experimental details are described in Ref. \onlinecite{huebner}. With this model, we are able to quantitatively compare the samples with different MAO barrier thicknesses. One drawback of the Brinkman model is its limitation to MTJ systems that do not show coherent tunneling. In addition, it is not able to explain features resulting from DOS related effects, such as half-metallic ferromagnetism. Thus, it is not possible to extract physically reasonable barrier parameters of the MTJs with MgO barrier, because of the coherent tunneling resulting from the $\Delta_1$ symmetry filter effect.

Figure \ref{fig:brink}(c) depicts the relative deviation of the calculated Brinkman barrier thickness (d$_ \text{B}$) from the nominal barrier thickness (d$_\text{N}$). With respect to the usual error range of the Brinkman model of 10\,\% (marked by the gray area), most of the elements are very close to the nominal barrier thickness. Except for the sample with an MAO barrier thickness of 1.4\,nm, this deviation does not exceed 15\,\%. An additional requirement of the Brinkman model to be applicable is d$_\text{N} > 1.0\,$nm. Apparently, the nominal barrier thickness of 1.4\,nm is too close to this limit, resulting in huge variations of the Brinkman barrier parameters. 

In general, the calculated Brinkman thicknesses depicted by the green squares in Fig. \ref{fig:brink}(c) are larger in the antiparallel (light) than in the parallel state (dark). For the barrier height and the barrier asymmetry in Figs. \ref{fig:brink}(d,e), we find a reversed behavior. Here, the parallel values (Fig. \ref{fig:brink}(d)) are generally larger than the antiparallel ones (Fig. \ref{fig:brink}(e)). Again, the results of the sample with a barrier thickness of 1.4\,nm MAO show a huge variation, while the results of all other samples are very consistent, even between different junction sizes. Excluding the results of the sample with an MAO barrier of 1.4\,nm, the barrier height decreases from 3\,eV to 1.7\,eV (1.9\,eV to 1.5\,eV) in the parallel (antiparallel) state. 

Additionally, the barrier asymmetry increases from $-1$\,eV to $0.5$\,eV in the parallel state, while it increases from -0.1\,eV to 0.4\,eV in the antiparallel state. Overall, the calculated values of the samples that have been prepared independently from the rest of the series (1.8\,nm and 2.0\,nm) show almost no deviation from the general trend in case of the barrier height. However, the values of the barrier asymmetry are different for the independently prepared samples. A possible explanation for this difference might be the deposition process, which plays a vital role for the barrier asymmetry.   

\subsection{Thermovoltages and Seebeck coefficients}\label{results3}

\begin{figure}[bt]\centering
		\includegraphics{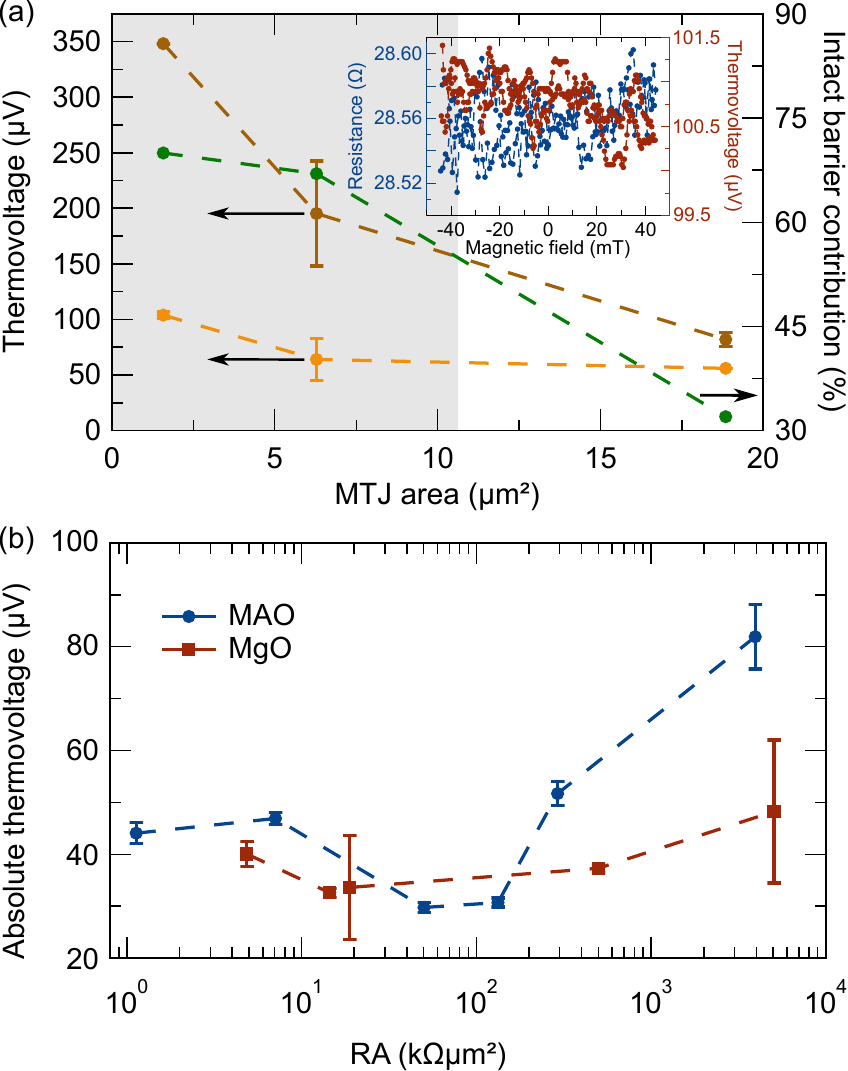}
	\caption{(a) Absolute thermovoltage (dark orange) and thermovoltage after dielectric breakdown (light orange) in dependence of the MTJ area. The inset shows TMR and TMS measurements after applying 3\,V to the junction and confirms the dielectric breakdown. Also, the contribution of the intact barrier to the absolute thermovoltage is shown (green) as well as the regime of homogenous heating (gray shaded area). (b) Measured absolute thermovoltages with a laser power of 150\,mW of the MTJs with an area of $6\pi$\,$\mu$m$^2$ in dependence of the RA product of all series.}
	\label{fig:thermovoltages}
\end{figure}

In order to investigate the high thermovoltages of the sample with MAO barrier, Fig. \ref{fig:thermovoltages}(a) depicts the thermovoltage in dependence of the MTJ area. Furthermore, the remaining thermovoltage after a dielectric breakdown of the junction is shown (see inset of Fig. \ref{fig:thermovoltages}(a)). Thus, it is possible to deduce the contribution of the intact tunneling barrier. The dielectric breakdown is confirmed via an additional TMR measurement after applying 3\,V to the junction. During the breakdown, the resistance changes from the M$\Omega$- to the $\Omega$-range. After the breakdown, both the TMR and TMS do not show any effect of magnetization switching (see inset of Fig. \ref{fig:thermovoltages}a).

Clearly, around 70\,\% of the absolute thermovoltage is caused by the intact tunneling barrier in case of small MTJ areas. This contribution decreases to 32\,\% for larger MTJ areas. Since the laser has a spot size of 11\,$\mu$m$^2$ when focused onto the MTJ \cite{TMS}, one possible explanation for the decrease is the occurrence of non-homogenous heating. Thus, additional lateral heat flows emerge, effectively lowering the temperature difference across the barrier and, ultimately, the measured thermovoltage as well. The regime of homogenous heating is indicated by the gray shaded area in Fig. \ref{fig:thermovoltages}(a), which represents the laser spot size. In future experiments, intermediate MTJ sizes could offer a more detailed insight into the processes of non-homogenous heating and in-plane temperature differences. With additional in-plane temperature differences, Nernst effects and additional magnetothermopower contributions become possible, which are not taken into account in current TMS experiments.

Figure \ref{fig:thermovoltages}(b) sums up the absolute thermovoltages in dependence of the RA product. Here, a laser power of 150\,mW is applied to MTJs with an area of 6$\pi$\,$\mu$m$^2$. In case of MAO, a drop of about 20\,$\mu$V in the absolute thermovoltage is measured for barrier thicknesses of 1.8\,nm and 2.0\,nm, which correspond to the samples that were prepared separately from the rest of the series. Since the RA products, the barrier heights and the Brinkman barrier thicknesses of series I are in good agreement with each other, the only difference is the barrier asymmetry. All other MTJs with MAO barriers show a thermovoltage that is consistently larger by a factor of up to 2 in comparison to the MTJs with MgO barrier.

In general, an increasing barrier thickness results in an increased temperature difference and, ultimately, in an increased measured thermovoltage. Thus, it is most likely that the contribution of the remaining stack to the absolute thermovoltage is different for the MTJs with MAO barrier thicknesses of 1.8\,nm and 2.0\,nm, for example via different lead contributions.
Excluding the two samples with MAO barriers, the difference between series I and series II is explainable by the different thermal conductivities of thin MAO and MgO films resulting in different temperature differences across the barrier and, thus, different thermovoltages.

Bulk MAO has a thermal conductivity of 23\,$\frac{\text{W}}{\text{K}\cdot \text{m}}$ \cite{mao2}, while bulk MgO has a thermal conductivity of 48\,$\frac{\text{W}}{\text{K}\cdot \text{m}}$ \cite{mgo}.  In Ref. \onlinecite{mgo} the thermal conductivity of thin MgO films is also experimentally determined to be 4\,$\frac{\text{W}}{\text{K}\cdot \text{m}}$. Taking the same reduction factor for thin MAO films, resulting in a thermal conductivity of 2.3\,$\frac{\text{W}}{\text{K}\cdot \text{m}}$, a COMSOL simulation offers insight into the actual temperature difference across the whole stack.

\begin{figure}[b]\centering
		\includegraphics{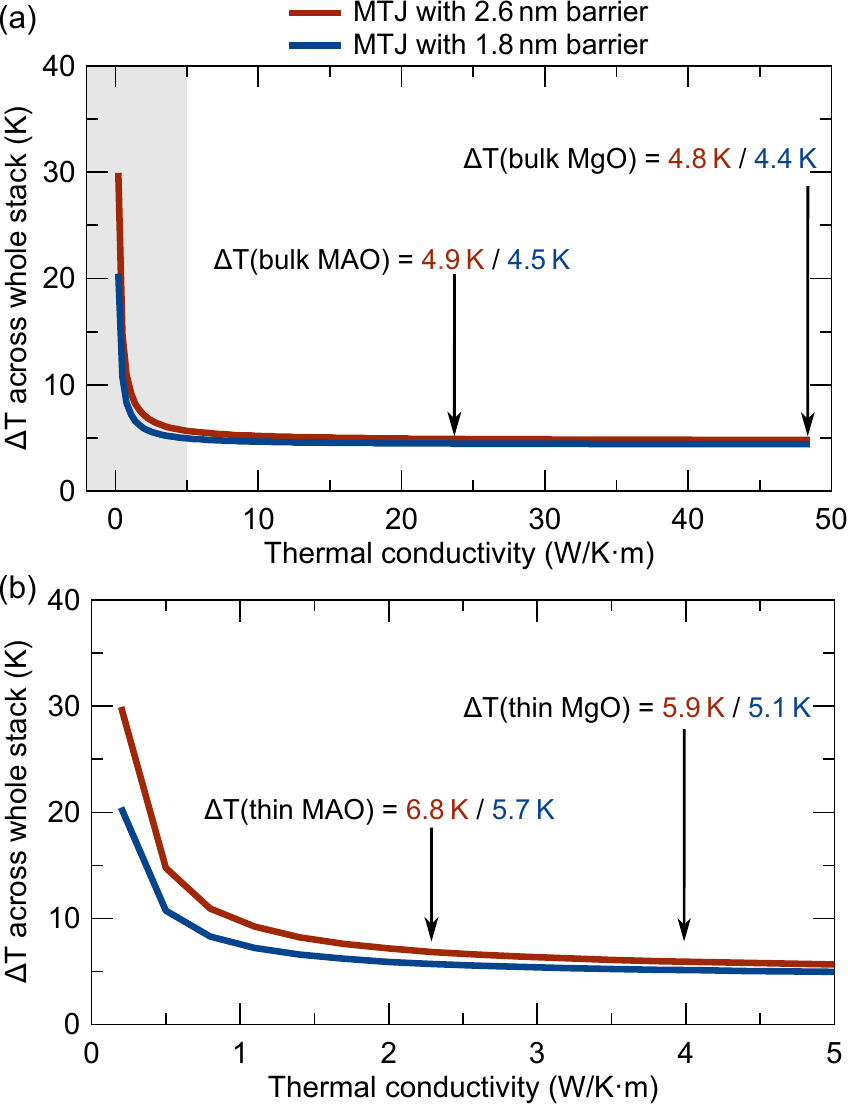}
	\caption{(a) Dependence of the temperature difference across the whole stack on the thermal conductivity of the barrier including both bulk values of MAO and MgO for a barrier thickness of 1.8\,nm and 2.6\,nm (laser power is 114\,mW, deduced from calibration measurements). The thin film regime is highlighted by the gray shaded area. (b) Thin film regime with both MAO and MgO values for the two barrier thicknesses.}
	\label{fig:COMSOL}
\end{figure}

Figure \ref{fig:COMSOL}(a) displays the result of this simulation for the interesting range of thermal conductivity and two barrier thicknesses. Accordingly, the thin film regime is shown in Fig. \ref{fig:COMSOL}(b). Since the area of the MTJs (1.6\,$\mu$m$^2$) is smaller than the area of the focused laser beam (11\,$\mu$m$^2$), the MTJs are heated homogeneously. The temperature differences become very large in comparison to values of preceding publications \cite{walter} (here, the laser spot area was usually around 240\,\,$\mu$m$^2$), since most of the laser beam energy is directly absorbed above the MTJ instead of a larger area of the Au bond pad. 
A systematic study of the influence of the laser spot size can be found in Ref. \onlinecite{TMS}. Of course, with the lack of actual measurements of the thermal conductivity of thin insulating films, COMSOL simulations offer only a limited insight into the actual thermal distribution inside an MTJ.

Hence, there is an ongoing discussion about the actual thermal conductivity of thin insulating films \cite{drop,drop2,drop3}. With the simulated temperature differences, the Seebeck coefficients for the MAO and the MgO MTJ with the highest TMS ratios are calculated (via $\text{TMS} = \frac{S_{\text{p}}-S_{\text{ap}}}{\text{min}(\left|S_{\text{p}}\right|,\left|S_{\text{ap}}\right|)}$) to be S$_\text{p}=-51\mu$V/K and S$_{\text{ap}}=-56\mu$V/K for MAO and S$_{\text{p}}=-12\mu$V/K and S$_{\text{ap}}=-15\mu$V/K for MgO, which is in good agreement with previous results \cite{walter,Boehnke2,Boehnke3,Boehnke1,huebner}.

\section{Conclusion}\label{conclusion}

We have studied the dependence of the laser-induced TMS effect on the barrier thickness of MAO and MgO MTJs and found a distinct maximum of the TMS ratio in case of thick barriers (nominal barrier thickness of 2.6\,nm) for both materials. The TMS ratio increased from (3 to 4)\,\% to 8\,\% for MTJs with MAO barrier, while the TMS ratio for MTJs with MgO barrier increased from around 15\,\% to 28\,\%. We found no experimental evidence of enhanced interface effects, which could explain the predicted increase of the TMS effect in case of thin barriers. The Brinkman model offered detailed insight into the barrier heights and asymmetries of the MTJs with MAO barrier. In addition, the extracted Brinkman barrier thicknesses provided a convenient way to compare samples with different nominal barrier thicknesses.

Furthermore, we measured very large thermovoltages of more than 350\,$\mu$V at the smallest MTJs of $0.5\pi$\,$\mu$m$^2$ with an MAO barrier, in contrast to 80\,$\mu$V for MTJs with a barrier of MgO. This difference is also reflected in the dependence of the thermovoltage on the barrier thickness. Here, MAO barriers show a thermovoltage that is larger by a factor of two in comparison to MgO barriers. Additionally, the MTJ with MAO barrier exhibits Seebeck coefficients that are thrice as large as for MTJs with MgO barrier (S$_{\text{p,MAO}}=-59\mu$V/K vs. S$_{\text{p,MgO}}=-18\mu$V/K) taking the reduced thermal conductivity of thin insulating films into account. 

Thus, we conclude that MAO is generally preferable as a barrier material when generating thermovoltages in MTJs. Still, further effort is needed to determine the real thermal conductivities of thin insulating films.

\section{Acknowledgments}

The authors gratefully acknowledge financial support from the Deutsche Forschungsgemeinschaft (DFG) within the priority program Spin Caloric Transport (SPP 1538).

\end{document}